\documentclass[a4paper,11pt]{article}
\usepackage{pos}

\usepackage{slashed}
\allowdisplaybreaks

\title{Gravitational $p \to \Delta^+ $ transition form factors in
  chiral perturbation theory}

\author*[a]{Bao-Dong Sun}

\affiliation[a]{Institut f\"ur Theoretische Physik II, Ruhr-Universit\"at Bochum,  D-44780 Bochum,
 Germany}

\emailAdd{baodong.sun@rub.de}

\abstract{The proton to $\Delta^+$ resonance transitional gravitational form factors are calculated to leading one-loop order using chiral perturbation theory in our recent work~\cite{Alharazin:2023zzc}. We take into account the leading electromagnetic and strong isospin-violating effects to obtain non-vanishing contributions. The loop contributions to the
transition form factors are found to be free of power-counting
violating pieces, which is consistent with the absence of
tree-level diagrams at the considered order. Our
results involve no free parameters and can be regarded as 
predictions of chiral perturbation theory.  
}

\FullConference{The 11th International Workshop on Chiral Dynamics (CD2024)\\
 26-30 August 2024\\
Ruhr University Bochum, Germany\\}

\begin{document}
\maketitle

\section{Introduction}
Hadrons including proton and neutron are building blocks of our universe. Nevertheless, their fundamental properties, such as mass, spin, and three-dimensional structures are still unclear in many aspects. These properties are closely related to the energy-momentum tensor (EMT) operator. 
The matrix elements of EMT can be written in terms of the gravitational form factors (GFFs)~\cite{Polyakov:2018zvc}.
At zero momentum transfer, these form factors correspond to the hadron mass, spin, and the so-called D-term which characterizes the distribution of forces inside the particles. Although the D-term is a property as fundamental as the mass and spin, it is much less understood. For a long time, it was believed that the GFFs can be accessed only through the graviton exchange, which is too weak to measure in any practical experiments. 

After the introduction of the generalized parton distributions (GPDs) in the '90s, it was soon realized that the GFFs are the second Mellin moments of the unpolarized GPDs~\cite{Ji:1996ek}. Since GPDs can be measured in the scattering experiments, their relations to the GFFs open a new window for the study of hadron structures. 
Unfortunately, due to the low statistics in the previous measurements, the extracted GFFs suffer from large uncertainties. For example, the measurement of the proton D-term by JLab data gives $-2.04\pm0.14(\text{stat.})\pm0.33(\text{syst.})$, which leads to a large error band in the calculation of the internal pressure distribution~\cite{Burkert:2018bqq}.  
To live with the large uncertainties, various effective approaches or models based on different assumptions of hadrons have been proposed in the past decades. These very different assumptions coexist within the large uncertainties of the experimental measurements. 

Beyond the scope of single particle, the transition processes are also interesting for the study of hadron structures. While the electromagnetic $p \to \Delta^+$ transition has been
  extensively studied over the past two decades on both the theoretical
  and experimental sides, see, e.g.,
  Refs.~\cite{Hilt:2017iup},
  the gravitational $p \to \Delta^+$ transition form factors (GTFFs)
  gained attention only since a few years \cite{Kim:2022bwn}.
  The GTFFs can also be accessed experimentally through their connection to the transition GPDs \cite{Frankfurt:1998jq,Frankfurt:1999xe,Diehl:2024bmd},
  obtained by expanding the non-local QCD operators 
  with various quantum numbers. 
Non-perturbative properties of the nucleon-$\Delta$ transition GPDs have been studied, e.g., by applying the approach of large $N_c$ limit of QCD, as discussed in Sec.~2.7 of 
Ref.~\cite{Goeke:2001tz}.  In Ref. \cite{Semenov-Tian-Shansky:2023bsy}, the transition GPDs have been connected with
  the DVCS amplitude within the process $e^- N \to e^- \gamma \pi N,$ while
  in Ref. \cite{Kroll:2022roq} these quantities have been studied using exclusive electroproduction
  of $\pi^- \Delta^{++}$. Recently, a complete definition of the $N \to \Delta$ transition GPDs has been provided in Ref.~\cite{Kim:2024hhd}.

     In Ref.~\cite{Kim:2022bwn}, the matrix element of the symmetric
     EMT corresponding to the $p \to \Delta^+$  transition has been
     studied for the first time, where a parametrization for the
     transitions $\frac{1}{2}^{\pm} \to \frac{3}{2}^{\pm} $ and
     $\frac{1}{2}^{\pm} \to \frac{3}{2}^{\mp} $ has been suggested in
     terms of five conserved and four non-conserved GTFFs. The first
     calculations of the GTFFs of the $N \to \Delta$ transition were
     done in Ref.~\cite{Ozdem:2022zig} using the QCD light-cone sum
     rules. The interpretation and understanding of the GTFFs have generated much
     interest recently, including the relation with the concept of QCD angular
     momentum (AM)~\cite{Kim:2023yhp}, etc.
 
For systematic studies of  low-energy hadronic processes involving the
$\Delta$ resonances and 
induced by gravity one may rely on the effective chiral Lagrangian for the nucleons, pions,
photons and delta resonances in  curved spacetime.
Effective Lagrangian of pions in curved spacetime has been obtained in Ref.~\cite{Donoghue:1991qv}, 
and the GFFs of the pion are considered in Ref.~\cite{Kubis:1999db}. 
The leading and subleading effective chiral Lagrangians for nucleons, delta resonances and
pions in curved spacetime,
along with the calculation of the leading one-loop 
contributions to the GFFs of the nucleons and the $\Delta$ resonances  can be found in
Refs.~\cite{Alharazin:2020yjv,Alharazin:2022wjj}\,\footnote{There are actually extensive applications of effective field theory grounded in chiral symmetry, e.g., Refs.~\cite{Wang:2019nvm,Meng:2021jnw,Wang:2019ato}.}. 

     In this proceeding we present our calculation in Ref.~\cite{Alharazin:2023zzc} of the GTFFs of the $p \to \Delta^+$
     transition in the
     framework of manifestly Lorentz-invariant chiral perturbation
     theory (ChPT) up-to-and-including the third order in the small-scale expansion
     \cite{Hemmert:1996xg}. Since the gravitational interaction respects the isospin
     symmetry, such kind of processes are possible
     only if the isospin symmetry is
     broken, i.e.~if $m_u \neq m_d$ and/or if the electromagnetic interaction is taken into
     account. We include both effects at the corresponding leading orders to calculate the
     one-loop contributions to the GTFFs.

This proceeding is organized as follows: In section~\ref{I}, we specify the
relevant terms of the effective Lagrangian in curved spacetime and expression for the parts of the EMT.   
We calculate the GTFFs of the $p \to \Delta^+$ transition in section~\ref{II}.  
The results of our calculations are summarized in
section~\ref{summary}. 

\section{Effective Lagrangian in curved spacetime and the energy-momentum tensor}
\label{I}

The action corresponding to the leading-order effective Lagrangian for
nucleons, pions, photons and delta resonances, interacting with an
external gravitational field, can be obtained from the
corresponding expressions in flat spacetime
\cite{Alharazin:2020yjv,Alharazin:2022wjj}.
It has the following form:
\begin{eqnarray}
S_{\rm \gamma}^{(2)}  
& = &
\int d^4x \sqrt{-g}\, \biggl\{\, -\frac{1}{4} F_{\mu \nu }F^{\mu \nu } + \frac{m_\gamma^2}{2} A_\mu A^\mu \biggl\},
\label{Gg}
\\
S_{\rm \pi}^{(2)} 
&=& \int d^4x \sqrt{-g}\, \left\{ \frac {F^2}{4}\,  {\rm Tr}
( D_\mu U  (D^\mu U)^\dagger ) + \frac{F^2}{4}\,{\rm Tr}(\chi U^\dagger +U \chi^\dagger) \right\}\,,
\label{PionAction} 
\\
S_{\rm N \pi}^{(1)}  
& = & 
\int d^4x \sqrt{-g}\, \biggl\{\, \bar\Psi \, i  \gamma^\mu
\overset{\leftrightarrow}{\nabla}_\mu \Psi -m \bar\Psi\Psi  +\frac{g_A}{2}\, \bar\Psi \gamma^\mu \gamma_5 u_\mu \Psi  \biggr\} \,, 
\label{PiNAction} 
\\ 
S_{\Delta \pi}^{(1)} 
& = &
 - \int d^4 x  \sqrt{-g} \biggl\{  
\Bar{\Psi}^{i \mu}  \,  i \gamma^\alpha \overset{\leftrightarrow}{\nabla}_\alpha  \Psi^{i}_\mu  -
m_\Delta \,  \Bar{\Psi}^{i}_\mu   \Psi^{i \mu}  -  g^{\lambda\sigma} \left( \Bar{\Psi}^{i}_\mu
i \gamma^{\mu}{\overset{\leftrightarrow}{\nabla}_\lambda}  \Psi^{i}_\sigma   +  \Bar{\Psi}^{i}_\lambda
i \gamma^{\mu}{\overset{\leftrightarrow}{\nabla}_\sigma}  \Psi^{i}_\mu  \right)  
\nonumber\\
&&+   i  \Bar{\Psi}^{i}_\mu \gamma^\mu \gamma^\alpha\gamma^\nu \overset{\leftrightarrow}{\nabla}_\alpha
\Psi^{i}_\nu + m_\Delta \Bar{\Psi}^{i}_\mu \gamma^\mu \gamma^\nu  \Psi^{i}_\nu +  \frac{g_1}{2}
\,g^{\mu\nu}\bar{\Psi}^i_{\mu} u_\alpha \gamma^\alpha \gamma_5 \Psi^i_{\nu}   
\nonumber \\
&&
+  \frac{g_2}{2} \bar{\Psi}^i_{\mu}  \left( u^\mu  \gamma^\nu + u^\nu \gamma^\mu \right) \gamma_5
\Psi^i_{\nu}   
+
\frac{g_3}{2} \bar{\Psi}^i_{\mu}   u_\alpha  \gamma^\mu \gamma^\alpha \gamma_5  \gamma^\nu \Psi^i_{\nu} 
\biggr\}\,,
\label{gdG}   
\\ 
S_{\Delta \rm N \pi}^{(1,2)} 
&=&
\int d^4 x \sqrt{-g} ~
\Biggl\{  - g_{\pi N \Delta} \bar \Psi \left( g^{\mu \nu} -\gamma^\mu\gamma^\nu\right) u_{\mu,i} \Psi_{\nu,i}  \nonumber\\
&&+
d_3^{(2)} i \bar \Psi f_+^{i \mu \nu} \gamma_5 \gamma_\mu \left( g_{\nu \lambda} -\left[z_n + \frac{1}{2}\right]  \gamma_\nu \gamma_\lambda \right) \Psi^{i \lambda}
+ \text{H.c.} \Biggl\}\,.
\label{gdnG} 
\end{eqnarray}
where $\Psi^\mu_i$ represent the $\Delta$ resonances, $\gamma_\mu \equiv e_\mu^a \gamma_a $ with $ e_\mu^a$ being the vielbein gravitational fields, $u$ is the matrix of pion fields, etc.
More details on the definitions for the building blocks of the effective Lagrangian and the choice of the external sources can be found in Ref.~\cite{Alharazin:2023zzc}.
The mass term ${m_\gamma^2} \, A_\mu A^\mu/2$ is introduced to
regularize possible infrared divergences (actually no IR divergences here), and the limit $m_\gamma \to 0$ should be performed  at the end. 

As gravity conserves isospin, the $p \to \Delta^+$ transition process exists only via the isospin-symmetry breaking effect, including strong interaction and electromagnetic interaction.
The leading isospin breaking effects due to the strong interaction is from the mass difference of the iso-multiplets: delta resonances, proton and neutron, and charged and neutral pions. To incorporate different masses, it's more connivent to work with fields in the physical basis than that in the isospin basis. The explicit relations for two basis is as follows:
\begin{eqnarray}
\Psi &=& \begin{pmatrix}
 \Psi_p  \\ \Psi_n \end{pmatrix}, 
 \Psi_{\mu,1} = \frac{1}{\sqrt{2}}
 \begin{pmatrix}
 \frac{1}{\sqrt{3}} \Delta^0_\mu - \Delta^{++}_\mu \\ \Delta^-_\mu  -  \frac{1}{\sqrt{3}} \Delta^+_\mu 
\end{pmatrix} , ~ 
 \Psi_{\mu,2} = -\frac{i}{\sqrt{2}}
 \begin{pmatrix}
 \frac{1}{\sqrt{3}} \Delta^0_\mu + \Delta^{++}_\mu \\ \Delta^-_\mu +  \frac{1}{\sqrt{3}} \Delta^+_\mu 
\end{pmatrix}, ~ 
 \Psi_{\mu,3} = \sqrt{\frac{2}{3}}
 \begin{pmatrix}
\Delta^+_\mu \\ \Delta^0_\mu 
\end{pmatrix}\,, \nonumber
\\
\pi^1& =& \frac{1}{\sqrt{2}} \left( \pi^+ + \pi^- \right),\, \pi^2 = \frac{i}{\sqrt{2}} \left( \pi^+ -\pi^- \right),\, \pi^3 =  \pi^0.
\end{eqnarray} 
After substituting the above definition of the fields into the action, e.g., Eq.~ (\ref{gdnG}), we get the terms relevant for the leading one-loop order contributions to the $p \to \Delta^+ $ transition:
\begin{eqnarray}
S_{\Delta \rm N \pi}^{(1)} 
&=& 
\int d^4x \sqrt{-g}\,   \frac{g_{\pi n \Delta}}{F} \Biggl\{ \bar \Psi_n \partial_\mu  \pi^+ O_2^{\mu \nu} \Delta^-_\nu  - \bar \Psi_p \partial_\mu  \pi^- O_2^{\mu \nu} \Delta^{++}_\nu  + \frac{1}{\sqrt{3}} \left(\sqrt{2} ~\bar \Psi_n \partial_\mu \pi^0 O_2^{\mu \nu} \Delta^0_\nu   \right.
\nonumber
\\
&&
 - \bar \Psi_n \partial_\mu \pi^- O_2^{\mu \nu} \Delta^+_\nu
+ \left. \bar \Psi_p \partial_\mu \pi^+ O_2^{\mu \nu} \Delta^0_\nu +  \sqrt{2}~ \bar \Psi_p \partial_\mu \pi^0 O_2^{\mu \nu} \Delta^+_\nu \right) \Biggl\}\,,
\end{eqnarray}
where $  O_2^{\mu \nu} = g^{\mu \nu} - \gamma^\mu \gamma^\nu$. 

The electromagnetic interaction also contributes to the mass splittings within iso-multiplets.
To obtain the leading isospin breaking effects due to the radiative corrections we do not distinguish between the masses of the isospin partners,
i.e. we take $m_{\Delta^{++}} = m_{\Delta^{+}} = m_{\Delta^{0}} = m_{\Delta^{-}}$, and $m_p = m_n$. The mass difference effect in electromagnetic interaction would be of higher order correction and therefore beyond the accuracy of our consideration.  Note that the separation of these contributions is afflicted with some uncertainties~\cite{Meissner:2022cbi}.

The EMT for bosonic matter fields can be obtained by varying the action $S$ coupled to a weak classical torsionless gravitational background field with respect to the metric $g_{\mu\nu}(x)$ according to
\begin{eqnarray}
T_{\mu\nu} (g,\psi) & = & \frac{2}{\sqrt{-g}}\frac{\delta S_{\rm m} }{\delta g^{\mu\nu}}\,.
\label{EMTMatter}
\end{eqnarray}
For example, from the action of Eqs.~(\ref{Gg}), we obtain the EMT terms in flat spacetime:
\begin{eqnarray}
T^{{(2)}}_{\gamma , \mu\nu} 
&=&
F_{\mu}^\alpha F_{\alpha \nu}+ m_\gamma^2 A_\mu A_\nu+\eta_{\mu \nu} \left( \frac{1}{4} F_{\alpha \beta }F^{\alpha \beta } -\frac{m_\gamma^2}{2} A_\alpha A^\alpha  \right),\,
\label{PhotonEMT}
\end{eqnarray}
where $\eta_{\mu\nu}$ is the Minkowski metric tensor with the
signature $(+,-,-,-)$.

For the fermionic fields interacting with the gravitational vielbein fields we use the definition
\cite{Birrell:1982ix} 
\begin{eqnarray}
T_{\mu\nu}  (g,\psi) & = & \frac{1}{2 e} \left[ \frac{\delta S }{\delta e^{a \mu}} \,e^{a}_\nu
+ \frac{\delta S }{\delta e^{a \nu}} \,e^{a}_\mu  \right]\,.
\label{EMTfermion}
\end{eqnarray}
For example, from the action of Eq.~(\ref{PiNAction}) we obtain the following terms for the EMT  in flat spacetime:
\begin{eqnarray}
T^{{  (1)}}_{\rm N, \mu\nu}  & = &  \frac{i}{2} \bar\Psi \,  \gamma_\mu  \overset{\leftrightarrow}{D}_\nu \Psi - \frac{\eta_{\mu\nu}}{2}\biggl(\, \bar\Psi \, i  \gamma^\alpha \overset{\leftrightarrow}{D}_\alpha \Psi -m \bar\Psi\Psi  \biggr) + \left( \mu \leftrightarrow \nu \right)   \,.
\label{MEMT}
\end{eqnarray}
The superscripts indicate the orders which are assigned to the corresponding terms of the action. The complete expressions of EMT  can be found in Ref.~\cite{Alharazin:2023zzc}.
    
 \section{Gravitational transition form factors to one loop}
\label{II}

The matrix element of the total EMT for the transition $p \to \Delta^+ $ can be
parameterized in terms of five form factors as follows \cite{Kim:2022bwn}:
\begin{eqnarray}
&&\langle  \Delta,p_f, s_f | T^{\mu\nu}  | N,p_i, s_i   \rangle 
\nonumber
\\
& = &  \bar u_\alpha(p_f, s_f) \Biggl\{ F_1(t) \left(  g^{\alpha \{\mu}P^{\nu\}} 
+ \frac{m_{\Delta^+}^2-m_p^2 }{\Delta^2} g^{\mu\nu} \Delta^\alpha - \frac{m_{\Delta^+}^2-m_p^2 }{2 \Delta^2} g^{\alpha\{ \mu}\Delta^{\nu\}} - \frac{1}{\Delta^2} P^{\{\mu} \Delta^{\nu\}} \Delta^{\alpha}  \right)  
\nonumber
\\
&+&
F_2(t) \left( P^{\mu}P^{\nu}\Delta^\alpha+ \frac{(m_{\Delta^+}^2-m_p^2 )^2}{4  \Delta^2} g^{\mu\nu} \Delta^\alpha -  \frac{m_{\Delta^+}^2-m_p^2 }{2 \Delta^2} P^{\{\mu} \Delta^{\nu\}} \Delta^{\alpha}  \right)  
\nonumber
\\
&+&
F_3(t) \left( \Delta^\mu  \Delta^\nu - \Delta^2 g^{\mu \nu} \right)  \Delta^\alpha   
\nonumber
\\
&+&
F_4(t) \left( g^{\alpha \{ \mu} \gamma^{\nu\} } +  \frac{2 (m_p + m_{\Delta^+})}{\Delta^2 } g^{\mu\nu} \Delta^\alpha -\frac{m_p + m_{\Delta^+}}{\Delta^2} g^{\alpha\{ \mu}\Delta^{\nu\}} - \frac{1}{\Delta^2}\gamma^{\{\mu} \Delta^{\nu\}} \Delta^{\alpha} \right)
\nonumber
\\
&+&
F_5(t) \left( P^{\{ \mu} \gamma^{\nu\} } \Delta^\alpha  +  \frac{(m_{\Delta^+}^2-m_p^2)  (m_p + m_{\Delta^+})}{ \Delta^2} g^{\mu\nu} \Delta^\alpha - \frac{m_p + m_{\Delta^+}}{ \Delta^2} P^{\{\mu} \Delta^{\nu\}} \Delta^{\alpha}  \right.
\nonumber
\\
&-&\left. \frac{m_{\Delta^+}^2-m_p^2}{2 \Delta^2} \gamma^{\{\mu} \Delta^{\nu\}} \Delta^{\alpha}  \right) \Biggl\} \gamma^5 u(p_i, s_i)\, ,
\end{eqnarray}
where $m_p$ and $m_{\Delta^+}$ are the proton and the $\Delta^+$ masses, respectively, $P = \left( p_f + p_i \right)/2$, $\Delta =  p_f - p_i $ and $t= \Delta^2$, and $P^{\{ \mu} \gamma^{\nu\} }=P^{ \mu} \gamma^{\nu }+P^{\nu} \gamma^{\mu} $.
As discussed before, the above matrix element is zero if the isospin symmetry is exact. 

To calculate the leading one-loop contributions to the above matrix elements,
one needs to organize different contributions according to a systematic
expansion, for which we employ the so-called $\epsilon$-counting scheme (also
referred to as the small scale
expansion) \cite{Hemmert:1996xg}\,\footnote{For an alternative power
  counting in ChPT with delta resonances see Ref.~\cite{Pascalutsa:2002pi}.}: the soft scale $Q$ is the order of the pion mass, and the chiral orders for the involving quantities are the following,
\begin{eqnarray} 
m_\pi & \sim & Q, \quad
m_{\pi^+}^2-m_{\pi^0}^2 \sim Q^2,  \quad
\text{$m_\Delta$ and $m_N$} \sim Q^0,  \quad
m_\Delta-m_N \sim Q^1,  \quad
m_p-m_n \sim Q^2,  \quad   
 \nonumber\\
\nabla_\mu \Psi_{(i)} & \sim & Q^0, \quad
 \text{$\mathcal{L}^{(N)}$ vertex} \sim Q^N,  \quad  
\text{electric charge } e \sim Q^1,  \quad 
\text{loop momenta} \sim Q^1,  \quad  
 \nonumber\\
 t &\sim& Q^2,  \quad
 \text{pion lines} \sim Q^{-2}, \quad
\text{$N$ and $\Delta$ lines} \sim Q^{-1}, \quad
\text{photon lines} \sim Q^{-2} ,
\label{power_counting}
\end{eqnarray}
where $\mathcal{L}^{(N)}$ is the interaction terms of the Lagrangian with chiral order $N$. The above power counting is realized after performing an appropriate renormalization to the manifestly Lorentz-invariant calculations, for which we choose the EOMS scheme of Refs.~\cite{Gegelia:1999gf}.

\begin{figure}[htbp]
 \begin{center} 
	\includegraphics[width=0.7\textwidth]{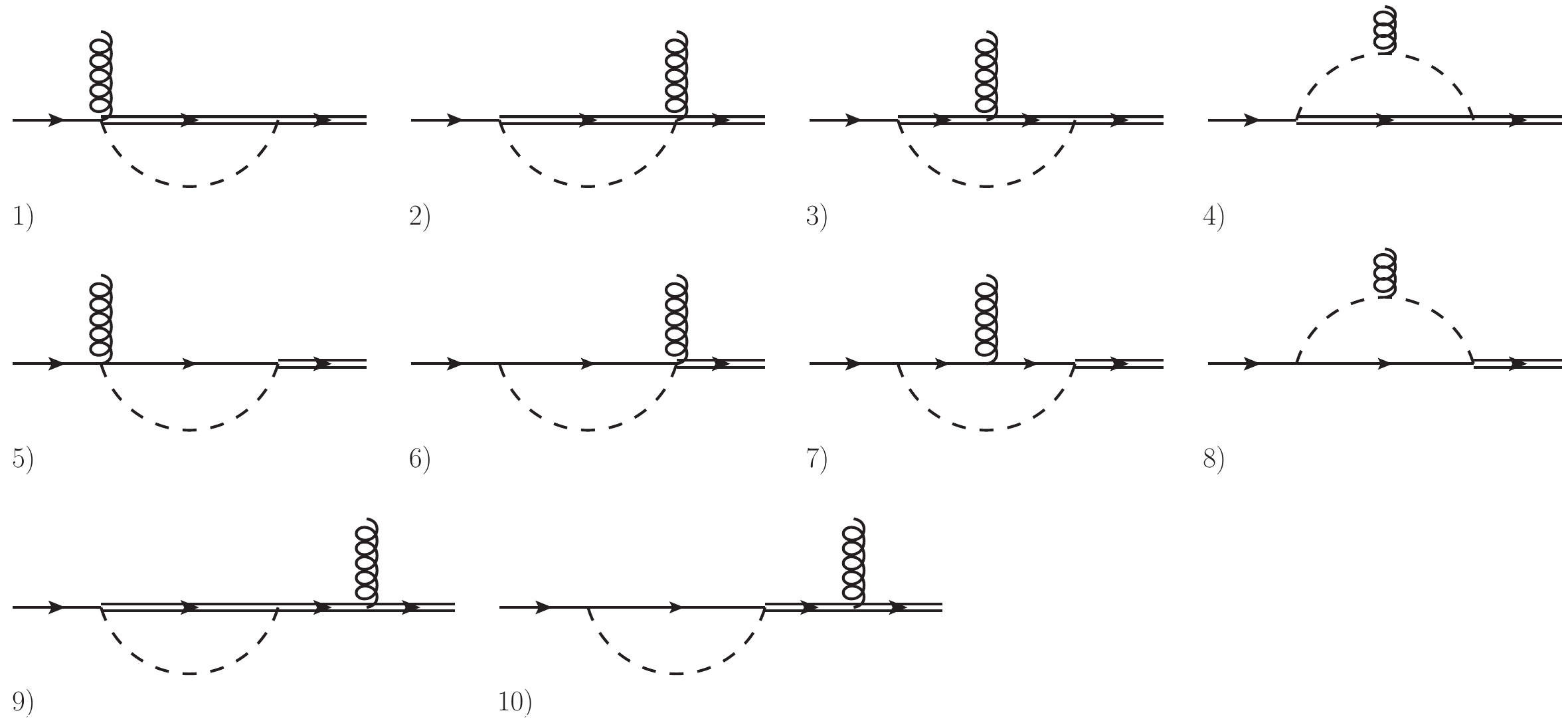}
	\caption{Strong contributions to the gravitational transition
          form factors.~Solid and double lines correspond to nucleons
          and $\Delta$ resonances, respectively. Dashed lines
          represent the pions, while the curly lines correspond to
          gravitons. Initial and final states refer to $p$ and
          $\Delta^+$, respectively, while the baryon lines inside
          loops refer to propagators of one of the following
          particles: $ \{\Delta^{++}, \Delta^+, \Delta^0, p,
          n \} $. Notice that the total contribution of these diagrams vanishes
          in the limit of exact isospin symmetry.
	\label{img:sp}}
      \end{center}    
\end{figure}

To obtain the one-loop contributions to the GTFFs due to strong
isospin-breaking interactions one has to compute 25 diagrams, where there are only 10
topologically differing diagrams and the rest can be obtained by just 
changing the masses and overall factors. These 10 diagrams are shown in Fig.~\ref{img:sp}, which give contributions of orders two and three.
We have verified that the result of diagrams in Fig.~\ref{img:sp} does not
contain power counting violating contributions and all ultraviolet divergences can be absorbed
into redefinition of the low-energy coupling constants of the effective Lagrangian.
The electromagnetic corrections can be obtained by replacing the pion line with the photon line in Fig.~\ref{img:sp}, which also does not
involve power-counting violating terms, and all ultraviolet
divergences can be absorbed as in the case of strong
isospin-breaking interactions. 

\begin{figure}[htbp]
	\centering
	\includegraphics[width=0.75 \textwidth]{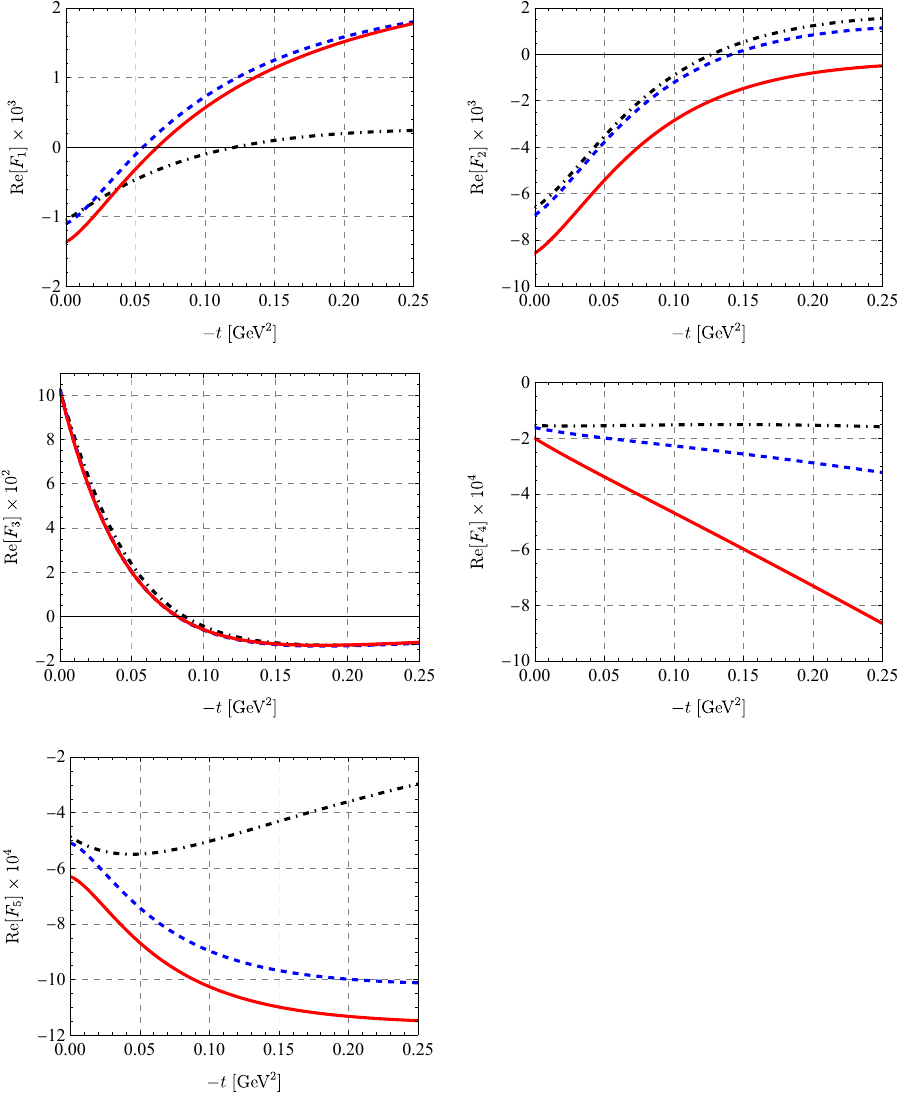}
	\caption{The real parts of the gravitational $p \to \Delta^+ $ transition
          form factors.~Dash-dotted (black),  dashed (blue) and solid 
          (red) lines correspond to the form factors containing
          contributions of loop diagrams with inner pion and nucleon lines
          only, diagrams with inner pion and nucleon lines plus radiative
          corrections, and all loop contributions,
          respectively.  }
	\label{imgR:em}
\end{figure}

For the numerical results, we used the following values of the involved
parameters:
\begin{eqnarray} 
 g_A & = & 1.289, \quad
g = 1.35,  \quad
m_{\pi^0} = 0.135,  \quad
m_p = 0.938,  \quad
m_n = 0.940,  \quad \nonumber\\
m_\Delta & =  &1.232,  \quad
F = 0.092,  \quad
m_{\pi^+}  =  0.140,  \quad
m_{\Delta^{++}} = 1.231 ,  \quad
m_{\Delta^+} = m_{\Delta},   \nonumber\\
m_{\Delta^0} & = & 1.233,  \quad
g_1 = 9 g_A/5, \quad 
e = 0.303 \,, \quad d_3^{(2)}=2.72\, {\rm GeV}^{-1}\,, 
\label{parameters}
\end{eqnarray}
where the various masses and the pion decay constant $F$ are given in GeV. 
In Fig.~\ref{imgR:em}, we present the numerical results of the obtained
strong and electromagnetic contributions to the real parts of the transition form
factors.  
There are also imaginary parts which are generated solely
by the loop contributions with internal nucleon lines and we do not show them here. 
The plots demonstrate that the diagrams with radiative corrections give smaller
contributions than the ones with pion loops in line with the power counting estimations.  
On the other hand the groups of diagrams with internal nucleon and delta lines give
comparable contributions. 

\begin{figure}[htbp]
	\centering
	\includegraphics[width=0.75 \textwidth]{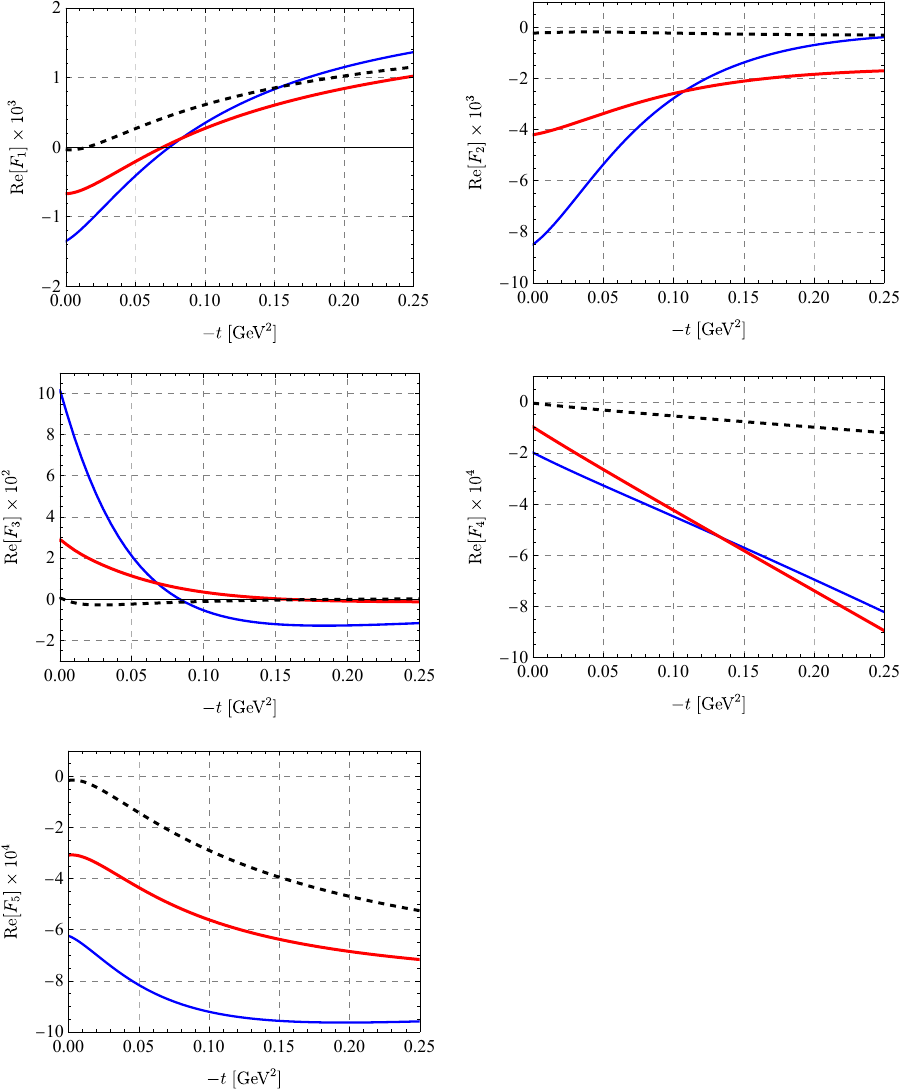}
	\caption{The real parts of the gravitational $p \to \Delta^+ $ transition
          form factors with same and different pion masses. The solid (blue), solid (red) and dashed (black) lines correspond to the form factors containing all loop contributions with charged pions having different mass than neutral pion, and that with same pion masses, and the radiative corrections only, respectively. }
	\label{imgR:pion}
\end{figure}

The leading contributions are given by terms proportional to the pion mass differences.
This is because these contributions involve integrals, whose integrands are proportional to
\begin{equation} 
\sim \frac{1}{p^2-m_{\pi^+}^2}- \frac{1}{p^2-m_{\pi^0}^2} \simeq  \frac{m_{\pi^+}^2-m_{\pi^0}^2}{\left(p^2-m_{\pi^+}^2\right) \left(p^2-m_{\pi^0}^2\right)}\,,
\label{diffPionPRs}
\end{equation}
and the contributions proportional to the proton-neutron mass difference (analogous for the case of the delta resonances)
are given by integrals, whose integrands are proportional to 
\begin{equation} 
\sim \frac{1}{\slashed p-m_p}- \frac{1}{\slashed p-m_n} \simeq \frac{m_p-m_n}{\left(\slashed p-m_p\right) \left(\slashed p-m_n\right)}\,.
\label{diffNuclPRs}
\end{equation}
According to the $\epsilon$-counting scheme in (\ref{power_counting}), the right-hand side of Eq.~(\ref{diffPionPRs}) has chiral order two, the same order as each of the terms in the left-hand side. In contrast, the right-hand side of Eq.~(\ref{diffNuclPRs}) has chiral order zero, one order higher than each of the terms on the left-hand side.
Therefore the total contribution of diagrams, which is proportional
to the proton-neutron mass difference, has the same order as the individual diagrams.  
While, the total contribution of diagrams, which is proportional
to the proton-neutron mass difference squared is suppressed by $Q$.
To show it more explicitly, we calculate the real parts of GTFFs with same and different pion masses and only the radiative corrections, given in Fig.~\ref{imgR:pion}. As one can see from the plots, for the absolute values of FFs at zero momentum, the case with different pion masses is at least twice as large as the case with same pion masses, and the radiative corrections are much smaller. The numerical result is what we would expect from above analysis. In region of momentum transfer squared that deviates from zero, the radiative corrections give
contributions of comparable sizes as the strong isospin violating effects.

\section{Conclusions and outlook} 
\label{summary}

In the framework of manifestly Lorentz-invariant ChPT for pions, nucleons, photons and
the delta resonances interacting with an external gravitational field, we calculated the leading
one-loop contributions to the gravitational transition form factors of the $p \to \Delta^+ $
transition process. 
	As the gravitational interaction respects the isospin
        symmetry, the amplitude of the $p \to \Delta^+ $ transition
        receives non-vanishing contributions due to isospin symmetry breaking. The results of
        the current work take into account the leading-order electromagnetic and strong
        isospin-breaking effects. 	
	Ultraviolet divergences and power counting violating pieces
        generated by loop diagrams in the manifestly Lorentz-invariant
        formulation of ChPT can be treated using the EOMS renormalization scheme.
However, at the order of our calculations, the one-loop contributions
to the form factors are found to be free of contributions that violate
the chiral power counting. This is consistent with the absence of
tree-level contributions at the considered order. For this reason, our
results involve no free parameters and can be regarded as 
predictions of ChPT. Notice, however, that the empirical information
on the mass splittings between the $\Delta$ resonance states, which enters as an
input in our calculations, is presently rather poor. 
Numerical results for the obtained transition form factors demonstrate
that the electromagnetic and strong isospin violating effects give
contributions of comparable sizes. This holds true for contributions
with both internal nucleon and delta lines.

\acknowledgments 

This work was supported in part by 
DFG and NSFC through funds provided to the Sino-German CRC 110
“Symmetries and the Emergence of Structure in QCD” (NSFC Grant
No. 11621131001, DFG Project-ID 196253076 - TRR 110),
by CAS through a President’s International Fellowship Initiative (PIFI)
(Grant No. 2018DM0034), by the VolkswagenStiftung
(Grant No. 93562), by the MKW NRW under the funding code NW21-024-A, by the EU Horizon 2020 research and
innovation programme (STRONG-2020, grant agreement No. 824093), by Guangdong Provincial funding with Grant
No. 2019QN01X172, the National Natural Science Foundation of China
with Grant No. 12035007 and No. 11947228, Guangdong Major Project of
Basic and Applied Basic Research No. 2020B0301030008, and the Department of Science and
Technology of Guangdong Province with Grant No. 2022A0505030010.

\end{document}